\documentclass[floatfix,pra,twocolumn,showpacs]{revtex4-1}

\usepackage{graphicx}
\usepackage{amssymb}
\usepackage{amsmath}
\usepackage{color}
\usepackage{psfrag}
\usepackage{epsfig}
\usepackage{bbm}
\usepackage{bm}
\usepackage{dsfont}

\usepackage{hyperref}
\hypersetup{breaklinks}

\newcommand{\ket}[1]{| #1 \rangle}

\newcommand{\rb}[1]{\left( #1 \right)}

\newcommand{\ew}[1]{\langle #1 \rangle}
\newcommand{\beq}{\begin{eqnarray}}
\newcommand{\eeq}{\end{eqnarray}}

\newcommand{\op}[2]{| #1 \rangle \langle #2 |}

\newcommand{\eq}[1]{Eq.~(\ref{#1})}
\newcommand{\fig}[1]{Fig.~\ref{#1}}

\newcommand{\bs}[1]{\boldsymbol{#1}}
\newcommand{\trace}[1]{\mathrm{Tr}\left\{#1\right\}}

\newcommand{\kett}[1]{| #1 \rangle\!\rangle }
\newcommand{\braa}[1]{\langle\!\langle #1|}
\newcommand{\eww}[1]{\langle\! \langle #1\rangle\! \rangle}

\begin{document}
\title{Decoherence and maximal violations of the Leggett-Garg inequality}
\author{Clive Emary} 
\affiliation{ 
  Institut f\"ur Theoretische Physik,
  Technische Universit\"at Berlin,
  D-10623 Berlin,
  Germany}

\date{\today}
\begin{abstract}
  We consider maximal violations of the Leggett-Garg inequality, obtained by maximising over all possible measurement operators, in relation to non-unitary aspects of the system dynamics.   We model the action of an environment on a qubit in terms of generic quantum channels and relate the maximal value of the Leggett-Garg correlator to the channel parameters.  We focus on unital channels, and hence on decoherence. In certain important cases, exact relations between the channel parameters and  maximal violations can be found.  
  Moreover, we demonstrate the existence of distinct thresholds for the channel parameters, below which no violation of the Leggett-Garg inequality can occur.
\end{abstract}
\maketitle

\section{Introduction}

The Leggett-Garg inequality (LGI) \cite{Leggett1985} has been the focus of much recent attention, both in theory \cite{Barbieri2009,Wilde2010,Avis2010,Lambert2010,Lambert2011,Montina2012,Wilde2012,Emary2012,Kofler2012,Emary2012a} and in experiment \cite{Palacios-Laloy2010,Goggin2011,Xu2011,Dressel2011,Waldherr2011,Athalye2011,Knee2012,Zhou2012}.
The inequality (or its simplest variant) reads
\beq
  K \equiv C_{21} + C_{32} - C_{31} \le 1
  \label{K3intro}
  ,
\eeq 
where $C_{\alpha\beta} = \ew{Q_\alpha(t_\alpha) Q_\beta(t_\beta)}$ is the
correlation function of the dichotomous variables $Q_{\alpha,\beta}=\pm 1$ at
times $t=t_\alpha$ and $t=t_\beta$.   Under the condition that the measurements are performed non-invasively, violation of this inequality implies the absence of a {\em macroscopic real} description of the system \cite{Leggett1985}.
Due to the existence of superposition states, quantum mechanics is not a macroscopic real theory, and can thus violate \eq{K3intro}. A two-level system undergoing Rabi oscillations, for example,  can achieve a value of $K = \frac{3}{2}$ for appropriate parameters.  Violation of the LGI by a quantum system is related to, and is typically taken as an indication of, quantum coherence in the system.

It is the aim of this paper to make a quantitative link between coherence --- or  rather, {\em decoherence}, the process by which coherence is lost --- and violations of LGI.
The point-of-view we take is that, if we had a perfect quantum system and could control its evolution as we please, then we should be able to violate \eq{K3intro} maximally. On the other hand, a system which supports no coherence (over the relevant time-scales) should certainly not produce a violation.  
In between these two extremes lies a spectrum of partially coherent dynamics and the questions is what happens for such intermediate cases?  What quantitative statements can we make concerning the degree and kind of decoherence and violations of LGI?

Our approach to answering these questions involves of two elements:  First, we describe the time evolution of the system between measurements in general terms by making use of the formalism of quantum channels \cite{King2001,Keyl2002}. Second, we shall study the {\em maximal} violations of \eq{K3intro} obtained by maximising over all possible choices of the measurement operators $Q_\alpha$.
This approach takes a significant part of the unitary component of the quantum evolution as given and allows us to connect $K_\mathrm{max}$, the maximised value of the correlator from \eq{K3intro}, to the non-unitary aspects of the dynamics.

The template for this approach is that taken with the (spatial) Bell's inequalities\cite{Bell1964,Bell2004}.  In the CHSH inequality\cite{Clauser1969}, for example, maximising over all possible detector angles establishes a direct relationship between the Bell correlator and the entanglement of the state under test (at least for pure state of two qubits) \cite{Gisin1991}.
For the CHSH  inequality, maximization over measurement angles connects the value of the Bell correlator with a property of the input state (entanglement).  In contrast, maximization over measurement angles for the LGI reveals the connection between the Leggett-Garg correlator, $K$, and a property of the {\em dynamics} (decoherence, dissipation).

Whilst our approach is general, we will focus here on a single qubit and environments whose effect on the system are Markovian.  We shall also focus on situations in which the system undergoes decoherence but not relaxation, i.~e. the channels we will mostly consider are unital channels \cite{King2001}.  Non-unital channels are briefly discussed in Sec.~\ref{SEC:nonU}

\section{LGI in terms of quantum channels}

We begin by describing our generic model of the measurement of \eq{K3intro} 
for a qubit.

\subsection{Measurements}
We assume that the measurements used to build $C_{\alpha\beta}$ are projections performed in an arbitrary basis.  Specifically, we parameterise the measurement of $Q_\alpha$ as first a rotation of the system with the unitary 
\beq
  q_\alpha 
  =
  \rb{
  \begin{array}{cc}
    \cos\rb{\frac{1}{2}\theta_\alpha} & e^{i \phi_\alpha}\sin\rb{\frac{1}{2}\theta_\alpha} \\
    -e^{-i\phi_\alpha}\sin\rb{\frac{1}{2}\theta_\alpha} & \cos\rb{\frac{1}{2}\theta_\alpha} 
  \end{array}
  }
  \label{measureq}
  ,
\eeq
followed by a projection on to $\sigma_z$ eigenstates, and then by the back-rotation, $q^\dag_\alpha$.
The measurement procedure can the be described by the map $\rho \to \mathcal{Q}_\alpha [\rho]$ acting on density matrix $\rho$.

\subsection{Quantum channels}
We will describe the time evolution between measurements in terms of the formalism of quantum channels \cite{King2001,Keyl2002}.
The most general map of one density matrix onto another, $\rho \to \Phi\left[\rho\right]$, can be written
\beq
  \Phi\left[\rho\right] = {v}^\dag S\left[u \rho u^\dag \right] v
  \label{Qchannel}
  ,
\eeq
where $u$ and $v$ are unitary rotations and $S$ is a superoperator responsible for non-unitary evolution whose action can be specified as follows.  Every density matrix can be written 
$
  \rho = \frac{1}{2}
  \rb{
    \mathds{1} +  \mathbf{w}.\bs{\sigma}
  }
$
where $\bs{\sigma}$ is the vector of Pauli matrices and $\mathbf{w}$ is a real, length-3 vector with $|\mathbf{w}|\le 1$.  Representing $\rho$ by length-4 vector $ \kett{\rho} =(1,\mathbf{w})$ the action of $S$ can be expressed in terms of a matrix acting on this vector: $S[\rho] \leftrightarrow S\kett{\rho}$.
The most general form of $S$ is
\beq
  S =
  \rb{
  \begin{array}{cccc}
    1 & 0 & 0 & 0 \\
    b_1 & c_1 & 0 & 0 \\
    b_2 & 0 & c_2 & 0 \\
    b_3 & 0 & 0 & c_3 \\
  \end{array}
  }
  ,
\eeq
with $b_i$ and $c_i$ real.  That the map must be stochastic (i.e. $S[\rho]$ is a valid density matrix) implies constraints on the coefficients $b_i$ and $c_i$.

The main focus of this work will be the so-called {\em unital} maps for which all $b_i$ are zero.  Channel $\Phi$ then maps the unit-matrix onto itself and describes a contraction of the Bloch sphere without displacement.  Such channels describe evolutions with decoherence but no relaxation.
In this unital case, the coefficients obey \cite{King2001}.
\beq
  |c_1 \pm c_2| \le |1\pm c_3|
  \label{unitalconstraint}
  .
\eeq

\subsection{LG correlation functions}

We assume that the time evolution of the system from $t_1$ to $t_2$ is described described by the map $\rho\to\Phi_1[\rho]$ and that from  $t_2$ to $t_3$ by $\rho\to\Phi_2[\rho]$.
This we assume to hold for the measurement of all three correlation functions $C_{ij}$ in \eq{K3intro}. In particular, the time evolution from $t_1$ to $t_3$ in the correlator $C_{31}$ is given by the composite $\rho\to\Phi_2[\Phi_1[\rho]]$.
This situation is not the most general, as it implies the absence of entanglement of the system with the bath at time $t_2$.  This model includes all Markovian environments.

With density matrix $\rho$ represented by vector $\kett{\rho}$, the action of each channel map and each measurement superoperator can be written as matrices acting from the left, e.g. $ \Phi_i[\rho]  \leftrightarrow\Phi_i\kett{\rho}$. In this representation, the trace operation is effected by multiplication from the left with the vector $\braa{\mathds{1}} = \frac{1}{2}(1,0,0,1)$. For generic superoperator $A$, we have
$
  \trace{A[\rho]} \leftrightarrow \eww{\mathds{1}|A|\rho}
$.
With measurements $ \mathcal{Q}_\alpha[\rho]  \leftrightarrow \mathcal{Q}_\alpha \kett{\rho}$, this notation allows us to write our expressions for the three LGI correlation functions as
\beq
  C_{21} &=& 
  \eww{\mathds{1} |
    \Phi_2 \mathcal{Q}_2 \Phi_1  \mathcal{Q}_1
  |\rho}
  \nonumber\\
  C_{32} &=& 
  \eww{\mathds{1} | 
    \mathcal{Q}_3 \Phi_2 \mathcal{Q}_2 \Phi_1
  |\rho}
  \nonumber\\
  C_{31} &=& 
 \eww{\mathds{1} |
\mathcal{Q}_3 \Phi_2  \Phi_1 \mathcal{Q}_1
  |\rho}
  .
\eeq

Since we are not explicitly interested in the unitary part of the evolution, we separate each $\Phi_i$ as in \eq{Qchannel} and redefine the measurement operators and initial state to remove the unitaries as far as possible.  After so doing, the only unitary part that remains can be expressed in terms the single rotation
$
  w=u_2 v_1^\dag
$.
With definition of the superoperator,
$
  W \kett{\rho} \leftrightarrow  w \rho w^\dag
$,
and its inverse,
$
  W^{-1} \kett{\rho}  \leftrightarrow  w^\dag \rho w
$, we obtain
\beq
  C_{21} &=& 
  \eww{\mathds{1} |
    \mathcal{Q}_2
    S_1
    \mathcal{Q}_1
  |\rho}
  \nonumber\\
  C_{32} &=& 
  \eww{\mathds{1} |
    \mathcal{Q}_3
    W^{-1}
    S_2
    W
    \mathcal{Q}_2
    S_1
  |\rho}
  \nonumber\\
  C_{31} &=& 
  \eww{\mathds{1} |
    \mathcal{Q}_3
    W^{-1}
    S_2
    W 
    S_1
    \mathcal{Q}_1
  |\rho}
  .
\eeq
Our task now is to maximise $K$ of \eq{K3intro} with these correlation functions over all measurement angles from the parameterisation \eq{measureq}. This task in too difficult to be performed analytically in the most general case.  We therefore consider some reduced cases of the greatest interest.  In all cases the maxima are found by setting the derivatives of $K$ with respect to the various angles to zero and solving.  Extensive comparison with numerical results was also carried out to ensure that the maxima described are global and not local.

\section{Single unital channel \label{SEC:1ch}}

We begin with a simple case, in which the second channel is purely unitary: $S_2 \to \mathds{1}$.  In this case, the rotation $W$ cancels from all correlators, which then involve only the measurements and the $S_1$ matrix. 
With $S_1$ unital, the three correlation functions in terms of the angles of \eq{measureq} read
\beq  
  C_{k1} 
  &=& 
  c_3 \cos\theta_1 \cos\theta_k
  \nonumber\\
  &&
  +\rb{\frac{}{}
    c_1 \cos\phi_1 \cos\phi_k
    +
    c_2 \sin\phi_1 \sin\phi_k
  }
  \sin\theta_1 \sin\theta_k;
  \nonumber\\
  C_{32} 
  &=& 
  \cos\theta_2 \cos\theta_3 + \cos\rb{\phi_2-\phi_3}\sin\theta_2 \sin\theta_3
  \label{1chC}
  , 
\eeq
for $k=2,3$.  These correlation functions, and thus the value of $K$, do not depend on the initial state $\rho$ (as has been observed for the LGI without dephasing \cite{Knee2012a}).

With these correlation functions, $K$ shows local maxima at values
$K = 1 + \textstyle{\frac{1}{2}} c_i^2$
\footnote{
  Example parameters that give this are:
    $\phi_\alpha = 0$,
    $\theta_1 = -\textstyle{\frac{1}{2}} \pi$,
    $-\sin \theta_2 = \sin \theta_3 = \textstyle{\frac{1}{2}} c_1$ 
 to obtain $K = 1 + \textstyle{\frac{1}{2}} c_1^2$;
    $\phi_1 = -\phi_2 = \phi_3 = \textstyle{\frac{1}{2}} \pi$,
    $\theta_1 = -\textstyle{\frac{1}{2}} \pi$,
    $\sin \theta_2 = \sin \theta_3 = \textstyle{\frac{1}{2}} c_2$
  to obtain $K = 1 + \textstyle{\frac{1}{2}} c_2^2$; and
    $\phi_1 = -\textstyle{\frac{1}{2}} \pi$,
    $\phi_2 = \phi_3$  ,
    $\theta_1 = 0$,
    $\cos \theta_2 = -\cos \theta_3 = \textstyle{\frac{1}{2}} c_3$
  to obtain $K = 1 + \textstyle{\frac{1}{2}} c_3^2$.
}.  
The maximal value for a given set of $c_i$ parameters is therefore
\beq
  K_\mathrm{max}^{(1)} = 1 + \textstyle{\frac{1}{2}} \mathrm{max}[c_i^2]
  \label{Kmax1ch}
  .
\eeq
Thus, providing that the channel can support some degree of coherence, i.e. the channel does not map every input state onto the maximally-mixed state, then some degree of violation of the LGI is possible since $K_\mathrm{max}^{(1)}>1$ provided at least one $c_i$ is non-zero.  The maximization over all measurement angles serves to pick out the direction in which $c_i$ is largest to obtain the largest violation.

\subsection{Example: Mach-Zehnder Interferometer}
The effect of decoherence on the test of LGI with Mach-Zehnder geometry \cite{Kofler2012,Emary2012a} can be formulated within this single-channel language.  If we assume that the decoherence of the two paths occurs exclusively before the second measurement, then the decoherence can be modelled by setting $c_3 = 1$ and $c_2=c_1$, with the value of $c_1$ depending on the strength of the decoherence.

According to \eq{Kmax1ch}, the maximal violation obtained with this set-up should be $K_\mathrm{max} = 3/2$ since $\mathrm{max} \left[c_i \right]= 1$.  This is the same as in the absence of decoherence.  In Ref.~\cite{Emary2012a}, however, decoherence was found to reduce the value of $K$ away from this value.  The resolution is that in Ref.~\cite{Emary2012a}, the measurement $\mathcal{Q}_2$ occurs in a fixed basis, namely the basis in which the off-diagonal elements of the density matrix are reduced by decoherence.  By fixing $\theta_2=0$ in our work, the three LG correlators read
\beq
  C_{21} &=& \cos \theta_1
  ;\quad
  C_{32}= \cos \theta_3
  ;\nonumber\\
  C_{31}&=&  \cos \theta_1 \cos \theta_3+ c_1 \cos\rb{\phi_1-\phi_3}\sin \theta_1 \sin \theta_3
\eeq
which (up to trivial redefinitions) are the same as found in Ref.~\cite{Emary2012a}.
Maximisation of $K$ over the remaining angles gives $K_\mathrm{max}=(1+|c_1|+c_1^2)/(1+|c_1|)$, which is equivalent that found in Ref.~\cite{Emary2012a}.
This shows that this simple single-channel model can reproduce the results of a proposed LGI test.

\section{Two unital channels}

We now consider the case where both channel matrices $S_i$ are the same.  Physically, this would most likely arise when the same decoherence mechanism is active throughout the experiment and the times in the LGI are equally spaced: $t_2 = t_1 +\tau$ and  $t_3 = t_1 +2\tau$, as is the most-often-studied case.

As in the foregoing, the specification of unital channels makes $K$ independent of the initial state.  We will study two cases separately here: first the case where the unitary component coming from the channels is absent: $W = \mathds{1}$ (which implies implies $u = v$); and then the case without this restriction.

\subsection{No $W$ rotation}

\begin{figure}[tb]
  \begin{center}
    \includegraphics[width=\columnwidth,clip]{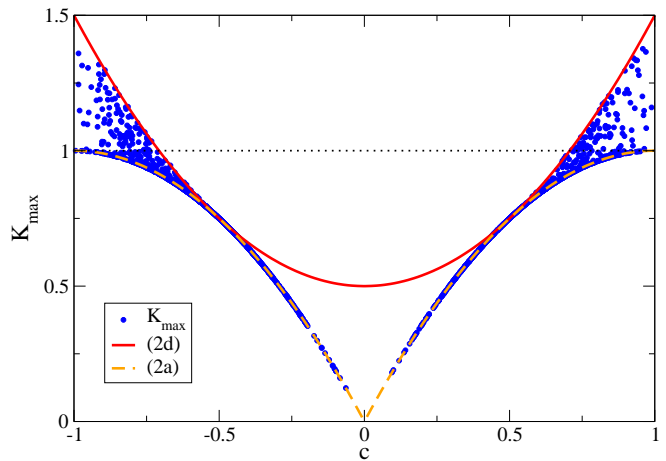}
  \caption{ 
    The maximised LG quantity $K_\mathrm{max}$ for the two-channel unital case with no additional rotations, $W = \mathds{1}$, plotted against the maximum decoherence parameter $c = \mathrm{max}\left\{c_i\right\}$.
    The blue points show $K_\mathrm{max}$ values calculated for a set of randomly generated set $c_i$ parameters.  The orange dashed line shows the lower bound $K_\mathrm{max}^{(2a)}$ and the red solid line shows the upper bound $K_\mathrm{max}^{(2d)}$.  Only for a decoherence parameter $c>1/\sqrt{2}$ can $K_\mathrm{max}$ exceed unity such that violations of the LGI may occur.
    \label{FIG:2ch_noW_unital}
 }
  \end{center}
\end{figure}

With $W = \mathds{1}$, the maximization of $K$ can be performed analytically.  Let us denote $c=\mathrm{max}\left\{c_i\right\}$, and $c'$ the next largest of the three $c_i$. Three maxima of $K$ are relevant:
\beq
  K^{(2a)}_\mathrm{max} =  |c|\rb{2 - |c|}
  \label{unital_noW_max_a}
  ,
\eeq
which is valid for all parameter values;
\beq  
  K^{(2b)}_\mathrm{max} = 1+ c^2\rb{1 - \frac{1}{c^2 + {c'}^2}} 
  \label{unital_noW_max_b}
  ,
\eeq
which is only valid when $\frac{|c'|}{c^2 + {c'}^2}\le 1$; and finally
\beq
  K^{(2c)}_\mathrm{max} =  \frac{c}{c^2 + {c'}^2} + {c'}^2
  \label{unital_noW_max_c}
  ,
\eeq
which is only valid for $\frac{|c|}{c^2 + {c'}^2}<1$
\footnote{
  If we assume that the coefficients are ordered $|c_1| \ge |c_3| \ge |c_2|$,
  $K^{(2a)}_\mathrm{max}$ can be obtained with $\theta_1 = \pi - \theta_3$,
  $\theta_2 = \pi/2$, and $\theta_3 = \pi c_1/2|c_1|$;
  $K^{(2b)}_\mathrm{max}$ can be obtained with $\cos\theta_1 = \frac{c_3}{c_1^2 + c_3^2} $,  $\theta_2 = 0$, and 
  $\theta_3 = -\theta_1$;
  and finally
  $K^{(2c)}_\mathrm{max}$ can be obtained with
  $\theta_1 = \pi - \theta_3$,
  $\theta_2 = \pi/2$, and 
  $\sin\theta_3 =  \frac{c_1}{c_1^2 + c_3^2}$.  In all cases we set$\phi_\alpha = 0$.
}.
Apart from the constraints mentioned, the value of $K_\mathrm{max}$ is the maximum of these three values.

Figure \ref{FIG:2ch_noW_unital} shows the value of $K_\mathrm{max}$ obtained for a set of randomly generated $c_i$ values, plotted as a function of $c$.  In generating the set of $c_i$ values, \eq{unitalconstraint} must be obeyed. 
For $|c|<1/2$ the maximal value depends only on the value $c$, with $K_\mathrm{max} = K_\mathrm{max}^{(2a)}\le 1$.  Above this value, the two further solutions come into play and, for a given value of $c$, $K_\mathrm{max}$ can take on a spread of values. The lower bound of this spread is simply $K_\mathrm{max}^{(2a)}$, the upper bound is found by taking $K_\mathrm{max}^{(2b)}$ with $c=c'$ which gives 
\beq
  K_\mathrm{max}^{(2d)} = \frac{1}{2}+c^2
  ,
\eeq
for $|c|>1/2$.

The violation of the LGI has a very clear threshold in this case --- the upper bound for $K_\mathrm{max}$ only exceeds unity if $c$ is greater that $\frac{1}{\sqrt{2}}$.  Thus, for violations of the LGI to occur, it is a necessary condition that
\beq
  c=\mathrm{max}\left\{{c_i}\right\} > \frac{1}{\sqrt{2}}
  \label{threshold}
  ,
\eeq  
and if dephasing is too strong, violation of the LGI can not occur.
\eq{threshold} is in itself not sufficient to ensure violation, as the lower bound lies within the classical bound for all values of $c$.  Consideration of the other parameters must be made to determine whether there is a violation or not.  One general remark can be made, however. If one or two of the $c_i$ are zero, then the maximum value of $K_\mathrm{max}$ is restricted to the solution $K^{(2a)}_\mathrm{max} \le 1$ and there can be no violation. This arises from the restrictions placed on the $c_i$ values by \eq{unitalconstraint}.

\subsection{With $W$-rotation}

\begin{figure}[tb]
  \begin{center}
\includegraphics[width=\columnwidth,clip]{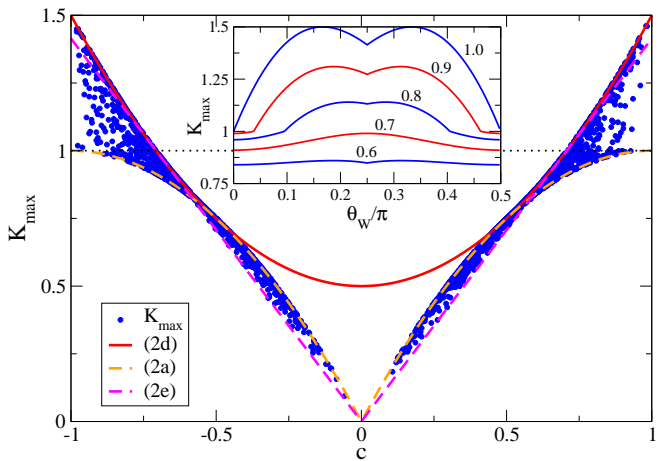}
  \caption{ 
    The same as \fig{FIG:2ch_noW_unital} but with a finite $W$ rotation.  In obtaining the blue data points, random values of the angles $\theta_W$ and $\phi_W$ are generated randomly alongside the decoherence parameters $c_i$  The upper bound (solid red line), and hence the threshold for LGI violations, is the same as in the $W=\mathds{1}$ case.   The lower bound is given by the minimum of either $K^{(2a)}_\mathrm{max}$ (orange dashed) or $K^{(2e)}_\mathrm{max}$ (pink dashed).
    INSET:  The two-unital-channel LG quantity $K_\mathrm{max}$ with $c_2=c_3=0$ and $\phi_W=0$ as a function of rotation angle $\theta_W$.  Shown are results for values of $c_1 = 0.6,0.7,0.8,0.9,1$, as marked. 
    \label{FIG:2ch_yesW_unital}
 }
  \end{center}
\end{figure}

The addition of the rotation $W$ complicates the maximisation of $K$ significantly, and an expression for the complete dependence of $K_\mathrm{max}$ on all parameters is not forthcoming.  Nevertheless, some exact statements can be made.
Figure \ref{FIG:2ch_yesW_unital} shows $K_\mathrm{max}$ for random values of the rotation angles $\theta_W$ and $\phi_W$ as well as $c_i$.
For $c<1/2$, the solution $K_\mathrm{max}^{(2a)}$ becomes the upper bound for $K_\mathrm{max}$, and a lower bound of
\beq
  K_\mathrm{max}^{(2e)}=\sqrt{2}c
 ,
\eeq
develops
\footnote{
  This lower bound can be obtained by setting e.g. $c_1=c_2=0$ and $c_3=c$ and angles $\phi_w=0$,  $\theta_1=0$, $\theta_2=\theta_W=\pi/4$ with the remaining angles arbitrary.
}.
For $c>1/2$, the upper and lower bounds on $K_\mathrm{max}$ are exactly the same as in the $W = 1$ case.
Importantly, this leaves the threshold behaviour unaltered: a $c$ of greater than $1/\sqrt{2}$ is required for violation of the LGI to be possible.

An illustrative, exactly-solvable, example of the dependence of the results on the relative-unitary angle $\theta_W$ obtains when $c_1$ is the only non-zero coefficient and $\phi_W =0$.  In this case, $K_\mathrm{max}$  is given by the maximum of
$
  |c_1| \rb{2 |\cos\theta_W| - |c_1|\cos 2 \theta_W}
$ 
or
$
  |c_1| \rb{2 |\sin\theta_W| + |c_1|\cos 2 \theta_W}
$.  
The behaviour of this maximum is illustrated in inset of \fig{FIG:2ch_yesW_unital} and shows a strong variation with angle $\theta_W$, even ranging from $1$ to $3/2$ in the limiting case of $c=1$.

\subsection{Example: Depolasing channel}

Let us consider that the channel in question is a depolarising channel, i.~e. unital with $c_i = c$.  In this case, we find
\beq
  K_\mathrm{max} = 
  \left\{
  \begin{array}{cc}
    |c|(1-|c|) & |c|\le 1/2 \\
    \frac{1}{2} + c^2 & |c|> 1/2 
  \end{array}
  \right.
  \label{Kmaxdepolar}
  ,
\eeq
irrespective of whether $W=\mathds{1}$ or not. In this case, $K_\mathrm{max}$ simply follows the upper bound in \fig{FIG:2ch_noW_unital}. That $K_\mathrm{max}$ is independent of $W$ as can be explained by the isotropy of $S$.

\subsection{Example: Damped Rabi oscillations}
The typical experimental system in which the LGI has to-date been investigated is a qubit undergoing damped Rabi oscillations. A complete channel description, including unitary part, of this dynamics, reads
\beq
  \Phi = 
  \rb{
  \begin{array}{cccc}
   1 & 0 & 0 & 0 \\
   0 & e^{-\Gamma t} &  0 & 0 \\\
   0 & 0 & e^{-\frac{1}{2}\Gamma t} \cos \Omega t 
     & e^{-\frac{1}{2}\Gamma t} \sin \Omega t \\
   0 & 0 & - e^{-\frac{1}{2}\Gamma t} \sin \Omega t
     & e^{-\frac{1}{2}\Gamma t} \cos \Omega t 
   \end{array}
  }
\eeq
where $\Omega$ is the Rabi frequency, $\Gamma$ the decoherence rate and $t$, the time.  This evolution corresponds to a unital $S$ matrix with $c_2 = c_3 = e^{-\frac{1}{2}\Gamma t}$ and $c_1 = e^{-\Gamma t}$, together with unitary rotation $W$ with angles $\phi_W=\pi/2$ and $\theta_W=\Omega t$.

Maximisation over all measurement angles gives the same result as for the depolarising channel, \eq{Kmaxdepolar}, independent of rotation angle $\theta_W$.  This follows because $W$ only acts in a subspace in which the two $c$-values are isotropic.
The threshold for violation of the LGI, \eq{threshold}, therefore holds exactly for this important case and, in terms of the original parameters, translates as $e^{-\frac{1}{2}\Gamma t} > 1/\sqrt{2}$.  With the time of maximal violation given by $\Omega t = \pi/6$ \cite{Leggett1985}, we obtain a condition on the ratio of decoherence rate and Rabi frequency
\beq
  \frac{\Gamma}{\Omega} < \frac{6}{\pi}\log 2 \approx 1.32
  ,
\eeq
for violations of the LGI to be possible.

\section{Non-unital channels\label{SEC:nonU}}

The general non-unital case is markedly more complex than the foregoing and we will not attempt to give a comprehensive analysis here. Nevertheless, it is worth highlighting a few important aspects that differ from the unital case.

\subsection{Single non-unital channel}

Repeating the single-channel calculation of section \ref{SEC:1ch} with a non-unital $S$-matrix, we observe --- and this is a general feature of the LGI for non-unital evolution --- that the value of $K$, and hence  $K_\mathrm{max}$, {\em does} depend on the initial state.  

In the single-channel case, however,  if the input state is the complete mixture $\rho = \mathds{1}$, it turns out that the LGI correlation functions are exactly the same as in \eq{1chC}.  This means that for this initial state, the non-unital results are independent of the parameters $b_i$.

In general, though, $K$ depends both on the initial state and on the non-unital parameters $b_i$.  By way of example, consider a ``classical channel'' \cite{King2001} with $b_1=b_2=c_1=c_2=0$ and $c_3=b_3=1/2 $, with pure state $\rho=\op{\psi}{\psi}$, $\ket{\psi} = \cos \Lambda \ket{\uparrow} + \sin \Lambda \ket{\downarrow}$ as input.  Maximisation over measurement angles gives $K_\mathrm{max} = \frac{1}{4}\rb{5+\cos2\Lambda}$, which oscillates between the classical bound ($K_\mathrm{max}=1$) and maximum violation ($K_\mathrm{max}=\frac{3}{2}$) as a function of input-state angle $\Lambda$.

\subsection{Two non-unital channels}

\begin{figure}[t]
  \begin{center}
    \includegraphics[width=\columnwidth,clip]{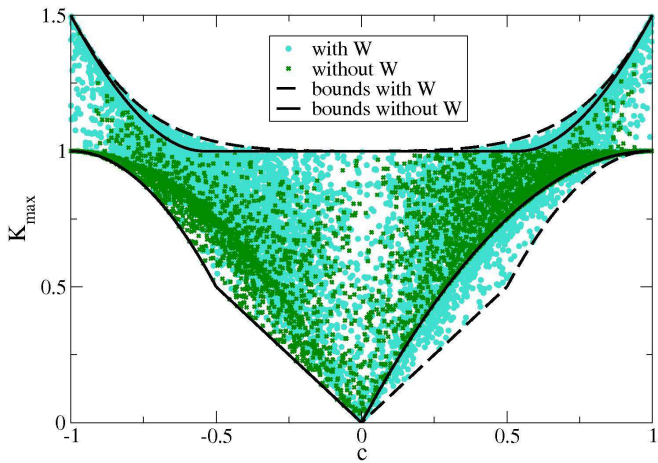}
  \caption{ 
    $K_\mathrm{max}$ in the non-unital case.  
    The green crosses show values for randomly generated channels with $W=\mathds{1}$; blue circles,  values for random parameters with finite $W$.
    The solid lines show the bounds without $W$ rotations; the dashed lines, the bounds without.  For $W=\mathds{1}$ violations of the LGI occur only when $c \gtrsim 0.544$.  For arbitrary $W$, no threshold is observed and violations are possible at all values of $c\ne 0$.
    \label{FIG:2ch_nonUnital}
 }
  \end{center}
\end{figure}
In the two-channel non-unital case, the LG parameter $K$ depends on the input state once again.   Here we shall only consider a completely mixed state as input.

Figure \ref{FIG:2ch_nonUnital} shows a set of $K$-values, together with the upper and lower bounds, for this non-unital situation. Results for both $W=\mathds{1}$ and with a finite-$W$ rotation are shown.
In the $W=\mathds{1}$ case, the bounds can be found analytically.  
The upper bound consists of two sections.  The first consists of the classical bound 
$K_\mathrm{max}=1$, which can be obtained by setting e.g. $c_1=c_2=b_1=b_2=0$; $c_3 = c$; $b_3 = 1-c$ and $\phi_\alpha=\theta_\alpha=0$.
The second can be found by setting $c_1=c_2 =c$, $b_1=b_2$, $c_3=b_3=0$ and then determining the value of $b_1$ as follows.  We consider the action of $S$ on pure state density matrix $\rho=n.\sigma$ with $n=(\cos A,\sin A \cos B,\sin A\sin B)$.  The length of the resulting Bloch vector, $l=\sqrt{
  \rb{b_1 + c \cos A}^2
  +
  \rb{b_1 + c \sin A \cos B}^2
}$, is maximal for input angles $(A,B) = (\pi/4,0)$. By setting $c+\sqrt{2}b_1= 1$, we obtain $l=1$ and it is with this condition that maximal value of $K$ is reached for a given value of $c$.   The expression for the upper bound so determined is
\begin{widetext}
\beq
  K_\mathrm{max}^{(3a)} &=& 
  \frac{1}{2 c}
   + \frac{1}{2} - \frac{c}{2} + \frac{c^2}{2} 
    - c^2 
    \cos\left\{
      \cos^{-1}\left[
        \frac{1 + c^2 + 3 c^4 - c^6}{4 (c^2 + c^4)}
      \right]
      + 
      \sin^{-1}\left[
      \frac{
        \sqrt{-1 - 2 c^2 + 9 c^4 + 28 c^6 
        + 9 c^8 + 6 c^{10} - c^{12}}
      }
      {
        2 c (1 + c^2)^2
      }
      \right]
    \right\}
  \nonumber
\eeq
\end{widetext}
The complete upper bound consists of the first value (unity) for $|c|\lesssim 0.544$, and $K_\mathrm{max}^{(3a)}$ above this value.
The lower bound shows an asymmetry with respect to the sign of $c$. For $c>0$, the bound is simply $K_\mathrm{max}^{(2a)}$. For $c<0$, it is
\beq
   K_\mathrm{max}^{(3b)} &=& 
 \mathrm{Max}
  \left[
    |c|,
    -1+ 4|c|-2c^2
  \right]
  \label{K3b}
  .
\eeq

With $W$ rotation, the bounds are extended somewhat.  The lower bound becomes $K_\mathrm{max}^{(3b)}$ of \eq{K3b} for all $c$.  The upper bound in \fig{FIG:2ch_nonUnital} could only be found numerically.  This was found by keeping just one $c_i$ non-zero, $c_1$ say, and then setting $b_3=0$ and $b_2 = \sqrt{1 - (c_1 + b_1)^2}$ such that the length $l=\sqrt{(c_1 + b_1)^2 + b_2^2+ b_3^2}$ is equal to one.  For a given $c_1$, we have to numerically optimise the value of $b_1$ that gives the maximum violation.  With the above set, the maximizing angles are $\theta_1 = \pi/2$, $\phi_\alpha = \phi_W =0$.  The remaining angles, $\theta_2$, $\theta_3$  and $\theta_W$, are then found numerically. 

The most significant feature of \fig{FIG:2ch_nonUnital} is that whilst the violation in the $W=\mathds{1}$ case still shows a threshold behaviour, with finite $W$ it does not. With $W=\mathds{1}$, violations are only possible for $|c|\gtrsim 0.544$ whereas violations, even if only small, are possible for all $c \ne 0$ in the finite $W$-rotation case.  
This enhanced violation arises because the relaxation present in the non-unital case can lead to an increased purity of the output state in comparison with that following from the same channel with $b_i=0$.

\section{Conclusions}

We have used here a maximisation over measurement angles to investigate the relationship between the LGI parameter $K$ and the non-unitary aspects of the dynamics as expressed through the quantum channel parameters $c_i$ and $b_i$.  The rotation $W$ (relative unitary) between the two-channels was also seen to play a role.

With just a single unital channel, the relation between $K_\mathrm{max}$ and the channel parameters is extremely simple, \eq{Kmax1ch}, and except for the trivial case in which the channel maps every state onto the unit matrix, violations of the LGI can always be found.  In contrast, the two-channel unital case shows a threshold behaviour, such that no violations occur if the maximum dephasing parameter $c$ is less than $1/\sqrt{2}$. This allows us to rule out violations of the LGI for a large class of environments.

The non-unital case is significantly more complex than the unital and we have only touched on it here. Importantly, the LG correlation function and its maximum value are no longer independent of the initial state. In the most general non-unital case, the threshold behaviour disappears, although this re-emerges for restricted classes of channel.
In the unital case, considering $K_\mathrm{max}$ as a function of $c$ provides a useful way of organising the behaviour of the LGI.  For the non-unital case, it is not clear whether this is still the case, since the effect on $K_\mathrm{max}$ of the displacements $b_i$ are on an equal footing to those caused by the contractions $c_i$.  More work is certainly needed on this front.

Our results set upper bounds on the possible values of $K$ for a given system. If in experiment the measurement basis is restricted, or equivalently if the unitary dynamics of the system is not completely controllable, then these additional restrictions may mean that the maximum value of $K$ obtained in experiment is less than the bound $K_\mathrm{max}$ here.  Knowledge of the maximal violation, however, may also suggest ways in which the experimental measurements could be modified to maximise the violation.
One further perspective on our results is that a measurement of $K_\mathrm{max}$ allows us to infer certain information about the channel(s) in question, in particular, bounds on $c $.  In this context, the local maxima of $K$, and not just its global maximum, can provide additional information, e.g. in the single channel unital case, knowing the local maxima provides all three $c_i$.  Whilst this is an interesting perspective on the violations of the LGI, it would not represent an efficient way to experimentally characterise the channels(s) in question.

Finally, we mention that the system-environment interaction considered here is not the most general, as we have excluded the possibility of system-bath coherence at the point where the second measurement is made.  The influence of strongly-coupled quantum environments on the LGI remains to be seen.

\begin{acknowledgments}
I am grateful to S.~Huelga and N.~Lambert for useful discussions.  This work was supported by the DFG through SFB 910.
\end{acknowledgments}


\end{document}